\documentclass[12pt,epsf,psfig]{article}
\input{epsf}
\usepackage[]{epsfig}  

\newcommand{\be}{\begin{equation}}
\newcommand{\ee}{\end{equation}}    

\newcommand{\mycaption}{\begin{center}%
Figure\thefigure%
\end{center}
\addtocounter{figure}{1}}

\begin{document}

\begin{center}

{\Large\bf Surface effects on the electronic energy loss of charged
particles entering a metal surface}\\ 

\vspace{0.5cm}

{A. Garc\'\i a-Lekue$^{\rm \,a,}$\footnotemark[1]
and
J.M. Pitarke$^{\rm \,a,b}$}\\

\vspace{0.6cm}

{\it $^{\rm a}$ Materia Kondentsatuaren Fisika Saila, Zientzi Fakultatea,
Euskal Herriko Unibertsitatea, 644 Posta kutxatila,
48080 Bilbo, Basque Country, Spain}\\
{\it $^{\rm b}$ Donostia International Physics Center (DIPC) and Centro Mixto
CSIC-UPV/EHU, Donostia, Basque Country, Spain 
}\\

\end{center}

\footnotetext[1]{ Electronic address: wmbgalea@lg.ehu.es; Fax:
+34-94-464-8500}

\begin{abstract}
Surface effects on the electronic energy loss of charged particles entering a
metal surface are investigated within linear-response theory, in the framework
of time-dependent density functional theory. Interesting phenomena occur in
the loss spectra originated by the boundary (bregenzung) effect, which is 
as  a consequence of the orthogonality of surface and bulk excitation
modes. Our calculations indicate that the presence of a non-abrupt
electron-density profile at the surface severely affects the nature of surface 
excitations, as deduced from comparison with simplified models. 
\end{abstract}



\section{Introduction}

A moving particle approaching a metal surface losses part of its
energy as a consequence of the creation of electron-hole pairs
and both bulk and surface collective excitations,  i.e., plasmons
\cite{Pines,Ritchie1}.  A theoretical description of these electronic
excitations is basic to understand the processes involved in several
spectroscopies,  such as x-ray photoelectron spectroscopy (XPS),
Auger-electron spectroscopy (AES), and reflection-electron energy-loss
spectroscopy (REELS)\cite{Raether}. Equally, the interaction of moving ions
with solids has represented an active field of basic and applied physics
\cite{arnau,prl}.

In this paper, we follow the theoretical framework developed in Ref.
\cite{Aran} and focus on the interaction of charged particles moving with
constant velocity along a definite trajectory that is perpendicular to a
jellium surface. Previous linear-response calculations of the energy loss of
charged particles entering a metal surface were carried out with the use of
simplified models \cite{Denton,Ritchie3}. Here we present  self-consistent
calculations of the energy-loss spectra, as obtained in the framework of
time-dependent density-functional theory (TDDFT) \cite{tddft}. Although in the
case of charged particles moving inside a solid nonlinear effects may be
crucial in the interpretation of energy-loss measurements \cite{pitarke1},
surface effects are expected to be well described within linear-response
theory, unless the velocity of the probe particle is small compared to the
Fermi velocity of the solid \cite{pitarke2}.

Contributions to the energy loss coming from the excitation of bulk and
surface plasmons are studied in detail, as well as the influence of an
appropriate characterization of the electron density at the surface.
 The
latter is found to be essential when the particle penetrates the
solid, as the
 surface region is particularly perturbed when the particle crosses the
surface.
 Besides, interference effects occurring between the various parts of
the ion trajectory are more relevant than in the case of charged particles
moving at a fixed distance from the surface, which claims for the need of a
self-consistent treatment of the surface electronic response.

In Section 2, we briefly discuss general expressions for the energy-loss
spectra of charged particles entering a solid surface along a trajectory
that is perpendicular to the surface. In Section 3, we report the results of
our full self-consistent calculations, and the effect that the electronic
selvage at a metal surface has on the energy-loss spectra is discussed by
comparing our full calculations with those obtained for electron densities
that drop abruptly to zero at the surface. Unless otherwise is stated, atomic
units are used throughout, i.e., $e^2=\hbar=m_e=1$. 

\section{Theory}

Let us consider a recoiless particle of charge $Z_1$ moving with
non-relativistic velocity ${\bf v}$ along a definite trajectory that is
perpendicular to a metal surface (see Fig. 1). The moving particle will be characterized
by a classical charge distribution. The solid will be described by a bounded
free-electron gas normal to the $z$ axis, consisting of a fixed uniform
positive background plus a neutralizing cloud of interacting electrons of
density $n(z)$.

Within linear-response theory, the probability per unit time and unit energy
for the probe particle at $z$ to transfer energy $\omega$ to the solid is
found to be \cite{Aran}
\begin{equation}\label{eq2} P_z(\omega)=-{Z_1^2\over
2\pi^2 v}\int_{-\infty}^{+\infty}dz'\,\cos[\omega(z-z')/v]\int dq\,q\,{\rm
Im}W(z,z';q,\omega),
\end{equation}
where ${\bf q}$ represents a wave vector parallel to the surface and
$W(z,z';q,\omega)$ is the so-called screened interaction 
\begin{equation}\label{screened2}
W(z,z';q,\omega)=v(z,z';q)+\int dz_1\int dz_2
v(z,z_1;q)\,\chi(z_1,z_2;q,\omega)\,v(z_2,z';q),
\end{equation}
$v(z,z';q)$ and $\chi(z,z';q,\omega)$ being two-dimensional Fourier
transforms of the bare Coulomb potential and the density-response function of
the solid \cite{Pines2}, respectively. The energy that the probe particle at
$z$ looses per unit path length due to the creation of electronic excitations in
the solid is simply \cite{pedro}
\begin{equation}\label{eq1}
-\frac{dE}{dz}=\frac{1}{v}\int_0^\infty d\omega\,\omega\,P_z(\omega).
\end{equation}

In the framework of TDDFT, the interacting density-response function
$\chi(z,z';q,\omega)$ is fully determined from the eigenfunctions and
eigenvalues of the Kohn-Sham equation of density-functional theory (DFT)
\cite{kohn65} and the exchange-correlation (xc) kernel $f_{xc}$ accounting for
short-range xc effects (see, e.g., Ref. \cite{Aran}), which in the
random-phase approximation (RPA) is taken to be zero. To compute 
$\chi(z,z';q,\omega)$, we first take a jellium slab of thickness 
$a  = 6\, \lambda_F$ 
\cite{note1} and electron density equal to the average electron density of
valence electrons in Al ($r_s=2.07$), and then solve the Kohn-Sham equation of
DFT in the local-density approximation (LDA) \cite{lda}, by following the
procedure described in Ref. \cite{Eguiluz}. For comparison, we also consider
simplified models for the screened interaction $W(z,z';q,\omega)$ of a
semi-infinite electron gas, which are derived for electron densities that drop
abruptly to zero at the surface: These are a hydrodynamic model (HDM)
\cite{Barton}, and a specular-reflexion model (SRM) \cite{Marusak} which
expresses the screened interaction in terms of the bulk dielectric function
$\epsilon(q,\omega)$. 

\section{Results}

First of all, we show results for the probablity $P_z(\omega)$ as a function
of $z$ and for selected values of the energy $\omega$, which we
obtain either in the full self-consistent surface RPA, in the HDM, or in the
SRM with the bulk RPA dielectric function. We set the velocity $v=4\,v_0$
($v_0=e^2/\hbar$ is the Bohr velocity) and $Z_1=\pm 1$, and our results can
then be used for arbitrary values of $Z_1$, as the energy-loss probability is
within linear-response theory proportional to $Z_1^2$. 

Fig. 2 shows the probability $P_z(\omega)$ for the moving particle
to transfer either the energy $\omega=\omega_p$ [$\omega_p=\sqrt{4\pi
n}$, $n$ being the average electron density, i.e., $1/n=(4/3)\pi r_s^3$], 
corresponding to the excitation of a bulk plasmon, or $\omega=\omega_s$ [
$\omega_s=\omega_p/\sqrt 2$], corresponding to the excitation of a surface
plasmon.  When $\omega=\omega_p$ [Fig. 2(a)], the energy-loss probability
increases as the projectile enters the surface and reaches a constant value
deep inside the solid. At the surface-plasmon energy [Fig. 2(b)], the
probability reaches a maximum when the probe particle is located near the
surface, and diminishes in the interior of the solid where only electron-hole
pairs can be excited with $\omega=\omega_s$.

It is well known that within a classical model consisting of a semi-infinite
medium of local dielectric function $\epsilon(\omega)$ the energy-loss
probability $P_z(\omega_s)$ would be maximum at $z=0$. However, the actual
dispersion of the surface plasmon, not included in the classical model, shifts
the peak position of $P_z(\omega_s)$ towards the interior of the
solid. Moreover, the so-called bregenzung or boundary effect reduces the
coupling to bulk plasmons by the presence of surface plasmons. As the
electronic selvage is changed from zero (HD and SRM) to its actual structure
(RPA), the creation of electron-hole pairs increases and the bregenzung effect
is more pronounced, thereby yielding larger surface-plasmon and smaller
bulk-plasmon excitation probabilities.

Fig. 3 exhibits our full surface RPA (solid line), HDM (dashed line), and SRM
(dashed-dotted line) calculations of the stopping power, as obtained from Eq.
(3) as a function of $z$ and with $v=4\,v_0$. As the projectile is entering the
surface, our full surface RPA energy-loss calculations show a broader and
smoother structure, which is mainly due to the presence of a non-abrupt
electron-density profile at the surface. Before entering the surface, the
selvage electronic structure yields a larger energy loss than
predicted within SRM. Inside the solid the actual energy loss is smaller
than in the SRM, due to the bregenzung effect arising from the orthogonality
of the surface and bulk plasmon modes. In the interior of the solid, where the
electron density is constant, both our full surface RPA and SRM calculations
coincide with the well-known RPA stopping power of a uniform electron gas
(horizontal dotted line).  On the other hand, the  HD calculation
 deep inside 
the solid coincides with the bulk HD, which differs from the bulk RPA.

In conclusion, our full self-consistent RPA calculations of the energy loss of
charged particles entering a metal surface indicate that a fully
quantum-mechanical treatment of the electronic response of a metal surface is
necessary to quantitatively describe the effect of the presence of the
surface. A systematic self-consistent investigation of both RPA and
beyond-RPA energy-loss spectra of charged particles moving along arbitrary
trajectories is now in progress.   

We acknowledge partial support by  the Basque Unibertsitate eta Ikerketa
Saila and the Spanish Ministerio de Educaci\'on y Cultura.

\clearpage

\begin{figure}[hbt!]\label{fig1}
\caption{Illustration of an external charged particle impinging perpendicularly on a
metal surface.}
\end{figure}

\begin{figure}[hbt!]\label{fig2}
\caption{Energy-loss probability $P_z(\omega)$ versus $z$ for a particle of
charge $Z_1=\pm 1$ moving with velocity $v=4\,v_0$ to transfer the energy
(a) $\omega=\omega_p$ or (b) $\omega=\omega_s$ to the solid, corresponding to
the excitation of a bulk and surface plasmon, respectively. The
electron-density parameter $r_s$ is taken to be that corresponding to the
average density of valence electrons in Al, i.e.,
$r_s=2.07$. The result of our full self-consistent RPA calculation is
represented by a solid line. The results obtained with the use of HD and SR
models are represented by dashed and dashed-dotted lines, respectively. }
\end{figure}

\begin{figure}[hbt!]\label{fig3}
\caption{Stopping power, as obtained from Eq. (3) as a function of $z$ and
with $v=4\,v_0$. As in Fig. 2, solid, dashed, and dashed-dotted lines
represent the result of our full self-consistent RPA calculations and those
obtained with the use of HD and SR models, respectively. The horizontal dotted
line represents the result of a bulk RPA calculation.}
\end{figure}

\clearpage

\setcounter{figure}{1}

\begin{figure}
\centering
\epsfig{file=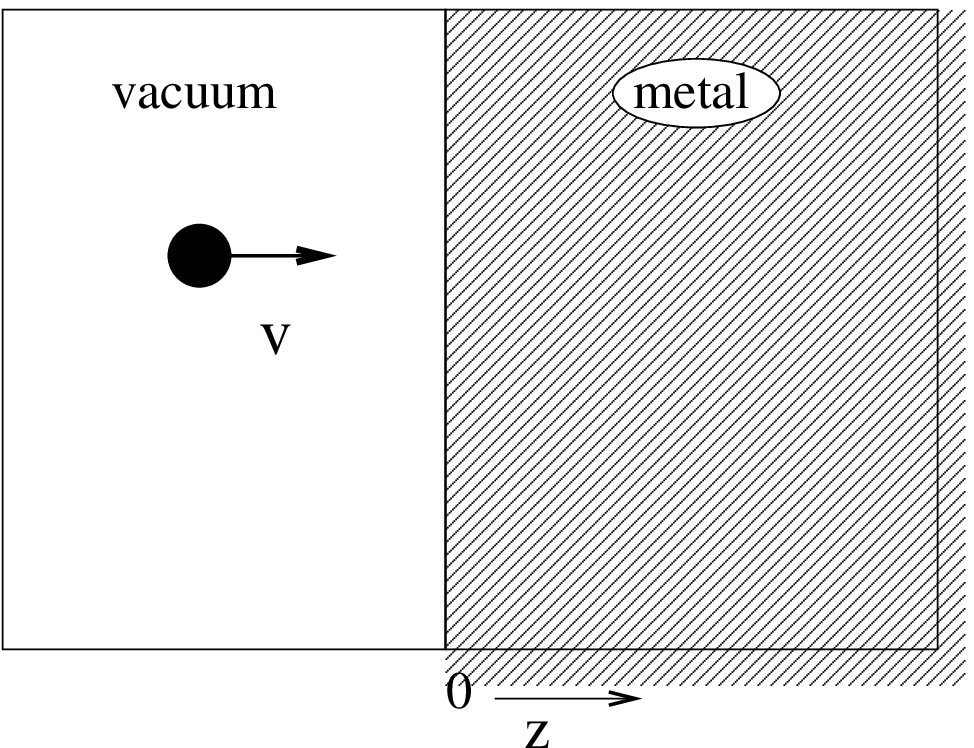, width=7cm, height=9cm,angle=0}
\mycaption
\end{figure}

\begin{figure}
\centering
\epsfig{file=fig2.eps, width=14cm, height=18.11cm,angle=0}
\mycaption
\end{figure}

\begin{figure}
\centering
\epsfig{file=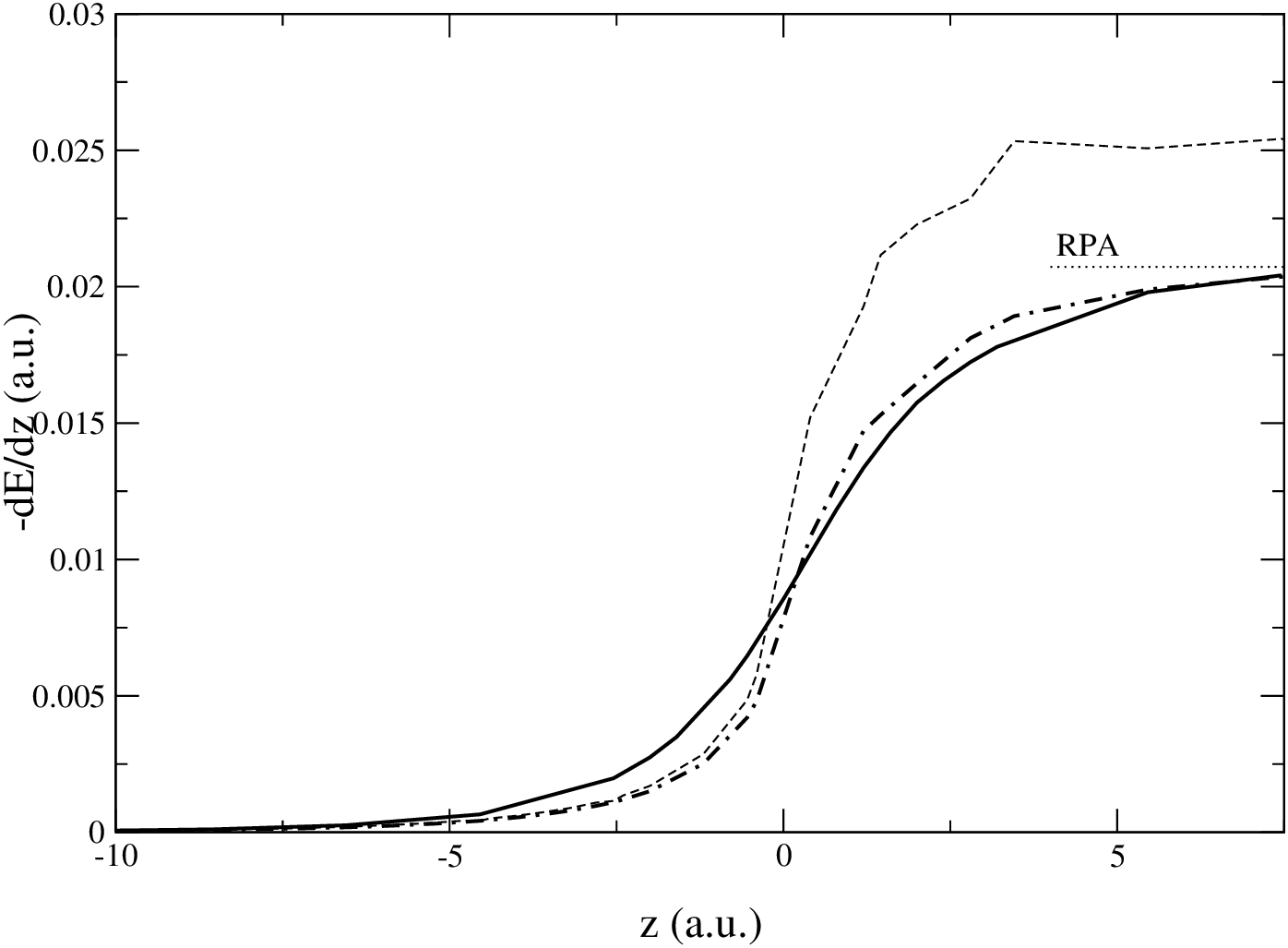, width=18.11cm, height=14cm,angle=90}
\mycaption
\end{figure}


\begin{thebibliography}{99}
\bibitem{note1} The results obtained for this slab thickness are
well-converged and are expected to accurately describe the case of a
semi-infinite free-electron gas. $\lambda_F = 2\pi r_s/(9\pi/4)^{1/3}$.
\end{thebibliography}
\end{document}